# Dandelion Diagram: Aggregating Positioning and Orientation Data in the Visualization of Classroom Proxemics

**Pengcheng An**
Eindhoven University of Technology
Eindhoven, the Netherlands
p.an@tue.nl

**Saskia Bakker**
Philips Experience Design
Eindhoven, the Netherlands
saskia.bakker@philips.com

**Sara Ordanovski**
**Chris L.E. Paffen**
Utrecht University
Utrecht, the Netherlands
c.l.e.paffen@uu.nl

**Ruurd Taconis**
**Berry Eggen**
Eindhoven University of Technology
Eindhoven, the Netherlands
r.taconis@tue.nl
j.h.eggen@tue.nl

## Abstract
In the past two years, an emerging body of HCI work has been focused on classroom proxemics—how teachers divide time and attention over students in the different regions of the classroom. Tracking and visualizing this implicit yet relevant dimension of teaching can benefit both research and teacher professionalization. Prior work has proved the value of depicting teachers' whereabouts. Yet a major opportunity remains in the design of new, synthesized visualizations that help researchers and practitioners to gain more insights in the vast tracking data. We present Dandelion Diagram, a synthesized heatmap technique that combines both teachers' positioning and orientation (heading) data, and affords richer representations in addition to whereabouts—For example, teachers' attention pattern (which directions they were attending to), and their mobility pattern (i.e., trajectories in the classroom). Utilizing various classroom data from a field study, this paper illustrates the design and utility of Dandelion Diagram.

## CSS Concepts
• **Human-centered computing~Visualization**

## Introduction
Recently, an emerging body of HCI research has been focused on tracking and visualizing *classroom proxemics* [1, 2, 16–18]—how teachers allocate time and attention to interact with students in different regions of the classroom during pedagogical activities. In educational science, the use of space in classroom pedagogy [15], or teacher-student physical proximity [12], has long been qualitatively recognized as a relevant yet implicit dimension of teaching that has major influences on learners [4, 8, 20, 21] (e.g., on their engagement, participation, or rapport with the teacher). The recent development of more accurate, less costly sensor technologies promises better understandings of this implicit dimension of pedagogy enabled by vast and various tracking data.

Existing HCI work has shown how tracking and visualizing classroom proxemics could meaningfully advance educational research as well as teacher professionalization [1, 16]. However, a major opportunity still remains—i.e., how to design new visualizations that enable researchers and practitioners to more intuitively gain insights in the vast and various tracking data [18].

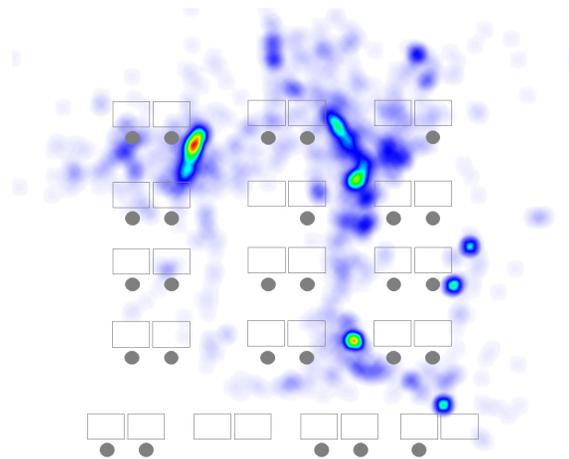

Regular heatmap based on positioning data (generated by the Pyheatmap library)

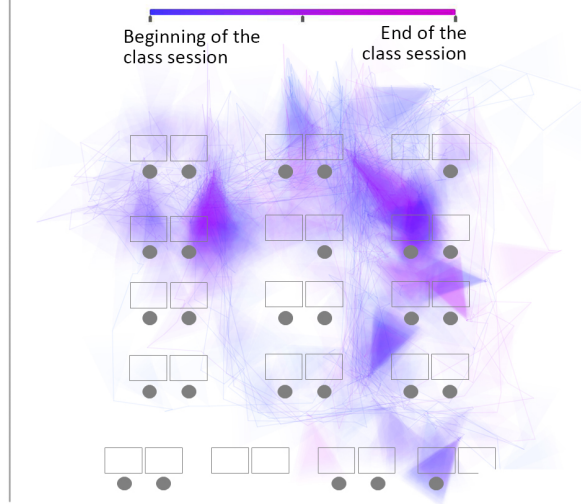

Dandelion Diagram based on positioning and orientation (heading) data

Beginning of the class session — End of the class session

**Figure 1**: Comparison between regular heatmap visualization based on solely teachers' positioning data (left), and Dandelion Diagram based on both teachers' positioning and orientation (heading) data (right). As shown above, Dandelion Diagram conveys richer insights in addition to whereabouts: e.g., teacher's attention pattern (i.e., which directions the teacher was attending to), mobility pattern (walking trajectories), and temporal pattern (i.e. during which period of the class session the teacher stayed at a location).

To tackle this opportunity, we present the design of Dandelion Diagram: a synthesized heatmap visualization which combines teachers' positioning and orientation (i.e., heading direction) data, to depict both their whereabouts and attention allocation in one frame (see Figure 1). Our design also enables the representation of teachers' mobility pattern (their trajectories in the classroom). Moreover, Dandelion Diagram contains an open-ended color coding layer that can be used to represent temporal pattern (i.e. during which period of the class session the teacher stayed at a location), or other information labels of the data points (e.g. which classroom activities the teacher was engaged in at a location).

In the remaining of this paper, we first briefly review the prior work on classroom proxemics and movement visual analytics. Subsequently, we introduce the rationale and design components of Dandelion Diagram, and illustrate its utility using data gathered in a prior field study [1] from diverse types of secondary school classrooms.

**Related Work**

*Classroom proxemics*
Fundamentally, *proxemics* functions as an important non-verbal channel in our daily social communication [9]. As its embodiment in classroom contexts, classroom proxemics mainly refers to the spatial aspects of teacher-student interactions [16]: such as how a teacher divides time and attention over students in the different regions of the classroom. Although classroom proxemics is an implicit dimension in teacher performance, many educational studies have suggested its major influences on learners, including their motivation, engagement, participation, and perceived closeness to the teacher [5, 12, 15]. And teachers are increasingly expected to develop skills of reflecting on classroom proxemics and using it to positively influence learners [2, 8, 20].

Echoing this, an emerging body of work in HCI has presented a few cases of tracking and visualizing classroom proxemics, to support both educational research and teacher professionalization [1, 2, 16, 17]. In the context of secondary education, for example, the work of ClassBeacons [1, 2] demonstrated how teacher proximity data can be presented in real time on ambient lamps in the classroom to enhance teachers' reflection-in-action. Addressing the context of university education, Martinez-Maldonado et al. [16, 17] foregrounded the value of classroom proxemics as a relevant aspect of multimodal analytics [17] to meaningfully complement existing approach to learning analytics. Moreover, they illustrated how a screen-based visualization of teacher positioning data could support teachers' post-hoc reflection [16].

While related work has showed the relevance of tracking and visualizing classroom proxemics, a major opportunity remains in the design of visualizations. Namely, more diverse explorations are still needed [18], in order to understand how different types of data can be meaningfully combined in visualization designs to enable richer and more intuitive insights in the vast tracking data. Towards this, our work explores a novel visualization design that combines both positioning data and orientation data, and affords richer insights in classroom proxemics in addition to teacher whereabouts.

*Movement visual analytics*
In the realm of movement visual analytics [3], a number of visualization designs have explored aggregating diverse types of data or representations in addition to location data (e.g. GPS or in-door positioning data) to enable new insights. For example, Andrienko et al. [3] introduced *Flower Diagrams,* a technique inspired by the renowned *Rose Diagrams* by Florence Nightingale. Flower Diagrams are used to represent spatiotemporal events at geographical scale: e.g. public transport stops in a city. The center of each "flower" represents a location where events happened (e.g. a bus station); and "petals" of the flower represent temporal change of the events (e.g. durations of the transport stops). The *Amoeba* visualization by Hyougo et al. [10] represents the movement distances and directions of people from a geographical location (e.g. a metro station).

At the scale of indoor spaces, heatmaps are often used to represent positioning data to analyze people's location patterns (e.g., see [7] and [13]). Meanwhile, related studies also explored other representations as alternatives or in addition to heatmaps, in order to, for example, study trajectory patterns of people in a shopping mall [22] or a museum [14]. However, orientation (e.g., heading) of people is rarely integrated in indoor movement analytics (see an example by Men et al. [19]), since existing research mostly concerns people's location, movement, or trajectory patterns rather than their attention pattern [6] (i.e. which directions they have been attended to).

By contrast, to understand classroom proxemics, teachers' orientation data can be as important as their positioning data [2]. With a regular heatmap, since the orientation cannot be seen, it is not clear whether a teacher *ignores* or, *over-attended* a specific part of the classroom. As Figure 1 shows, for a 'hotspot' location, orientation data can further reveal which one of adjacent students the teacher was directly interacting with. Our work therefore intends to meaningfully integrate orientation data in a synthesized heatmap.

**The Design of Dandelion Diagram**

In this section we detail the Dandelion Diagram design, including its rationales and design components.

*Which types of data are suitable*

Dandelion Diagram is designed for visualizing indoor tracking data of teachers, which are gathered continually between equal time intervals (e.g. every second). Each unit of data should at least contain the teacher's location (e.g., in Cartesian coordinate X, Y), and orientation (heading direction e.g., gathered by magnetic field sensor in radians or degrees). For instance, the data set used in this paper to demonstrate Dandelion Diagram are from the field study of ClassBeacons [1]. These data were gathered in every two seconds, through a wearable unit attached to the upper back of teachers (the orientation from upper back is likely to indicate their attention direction [2]). The wearable uses a commercial ultra-wideband (UWB) sensing kit [11]. Giving its accuracy (at decimeter level), UWB solutions has been opted for classroom proxemics tracking in multiple studies (e.g., [16]); however, other positioning tracking solutions could also be explored in future (e.g. computer vision).

*Which types of information to communicate*

In general, Dandelion Diagram is designed for communicating insights in following relevant aspects of classroom proxemics (also see Figure 1): *Location pattern*—where the teacher has stayed and for how long; *Attention pattern*—which directions the teacher was attending to at a certain location; *Mobility*

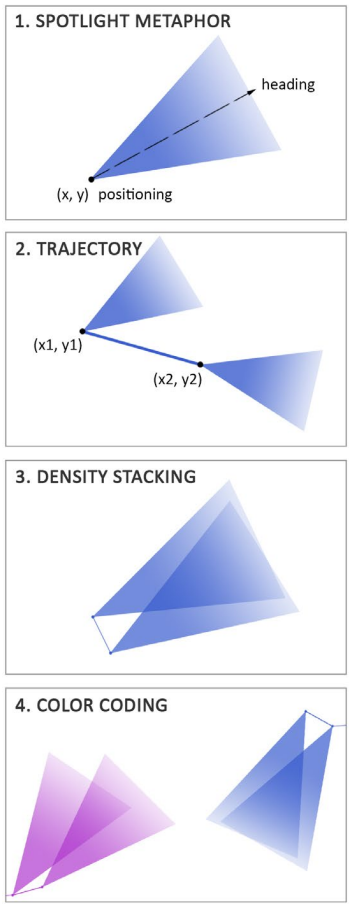

**Figure 1**: Illustrations of four major design components that constitute Dandelion Diagram.

*pattern*—the teacher's walking trajectories in the classroom; *Temporal pattern*—during which periods in the class session the teacher stayed at a location.

Location pattern (teachers' whereabouts) has been proved to be valuable for teachers' reflection [16]. On top of that, attention pattern gives further insights in teachers' behaviors [2]: e.g. in distinguishing which of the adjacent students the teacher was directly interacting with. Besides, teachers' presence can have different meanings to adjacent students: e.g. to students who are within teachers' angle of view and those who are not [2]. Mobility pattern, in addition, can depict teachers' wandering—which can have different interpersonal effects on learners than standing still and offering help [2, 16]. Finally, the temporal pattern can yield insights in teachers' proximity distribution over time (in addition to their proximity distribution over space), which proved to be relevant as well for teachers' reflection [16]. Now we further explain how Dandelion Diagram affords the mentioned insights, via a breakdown of its *four major design components*.

*Design component 1: **Spotlight metaphor***
As Figure 2(1) illustrates, using a *"spotlight"* metaphor, each data point of Dandelion Diagram is represented by a triangular shape, where the farthest vertex indicates the teacher's location, and its opposite side indicates the teacher's heading direction. Such *spotlight* representation has been widely used, e.g., in video games, to intuitively show the heading direction, as well as the position, of a person or an object.

*Design component 2: **Trajectory***
As Figure 2(2) illustrates, the vertices of every two consecutive *spotlight* units (i.e., every two consecutive position coordinates) are connected by a line. As such, the teacher's movement trajectories over the classroom can be delineated on the map.

*Design component 3: **Density stacking***
As shown by Figure 2(3), each spotlight unit is in semi-transparency, so that they can stack on one another to depict teachers' accumulated staying at a location over time through the accumulation of color density.

*Design component 4: **Color coding***
As Figure 2(4) illustrates, the color of each spotlight unit is mapped onto a continuous color spectrum, which represents the different time periods in a class session. However, this color coding feature could also be used to represent other labelled information of the data. For instance, multiple discrete colors could be used to represent different classroom activities (as used in [16]), or the data from different teachers, if there are multiple teachers in a class session. Consequently, the four design components jointly enable a Dandelion Diagram to communicate the relevant insights in classroom proxemics (see Figure 3).

## Implications for future work
In this section we discuss implications resulted from our design exploration of Dandelion diagram, which is aimed for informing future related work.

*Informal evaluation*
The design of Dandelion Diagram was informally evaluated with a teacher who had participated in the ClassBeacons study [1]. We visualized the data gathered from one of her class sessions during the study, and showed the visualization to her after her participation of the study. We did not introduce the

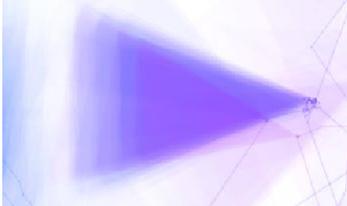
1. SPOTLIGHT METAPHOR

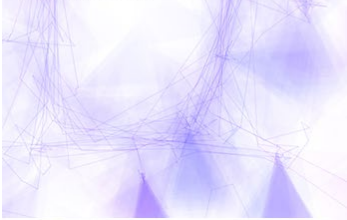
2. TRAJECTORY

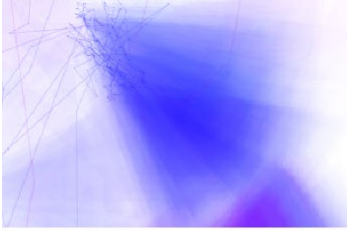
3. DENSITY STACKING

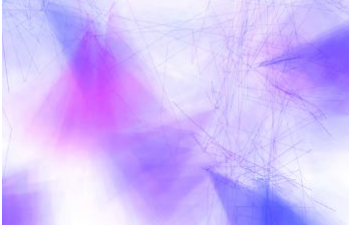
4. COLOR CODING

**Figure 3**: Examples of the four design components embodied by real-world classroom data.

design to her (we only explained the meaning of color coding), and asked her to openly describe and comment on the visualization, in order to see if the visualization could be intuitive to understand, and if it could support her reflection on that particular session. She reported that the visualization was easy to understand: "*It's immediately clear where I was*". Moreover, although the class session took place more than a week ago from the moment she saw the visualization, she could quickly recognize which lesson of the day it was and which students she was interacting with: "*I think this is the third class because […] so in front of the board, Elsa [alias], she had a lot of questions […] and this is Tony and Mark [aliases], they have a lot of questions […]*"

While this informal evaluation may imply the usability of Dandelion Diagram, better structured and more in-depth evaluation is clearly needed in the future, to further investigate how this visualization design could influence teachers' post-hoc reflection on their class sessions, and how this may influence their future practice.

*How classroom layout may shape classroom proxemics*
Besides how Dandelion Diagram might help teachers, we also learnt some implications in our exploration, which can be interesting to educational researchers. As shown by Figure 4, since the data set we use covers diverse types of classrooms [1], we are able to see a general difference of proxemics patterns between the "lecture-based" classrooms where the desks are placed in a matrix layout, and the "teamwork-based" classrooms where the desks are placed in irregular clusters. While we can generally see that the teachers' location, attention, mobility patterns appear differently in the two types of classrooms, it is not yet clear how these differences may influence teaching or learning, and how we could optimize the space of classrooms to fit different instructional or learning goals.

## Conclusion

We present Dandelion Diagram, a synthesized heatmap technique to visualize classroom proxemics. The technique combines both positioning and orientation data and affords richer insights in addition to whereabouts. Besides illustrating its design and utility, we also address the implications for future work.

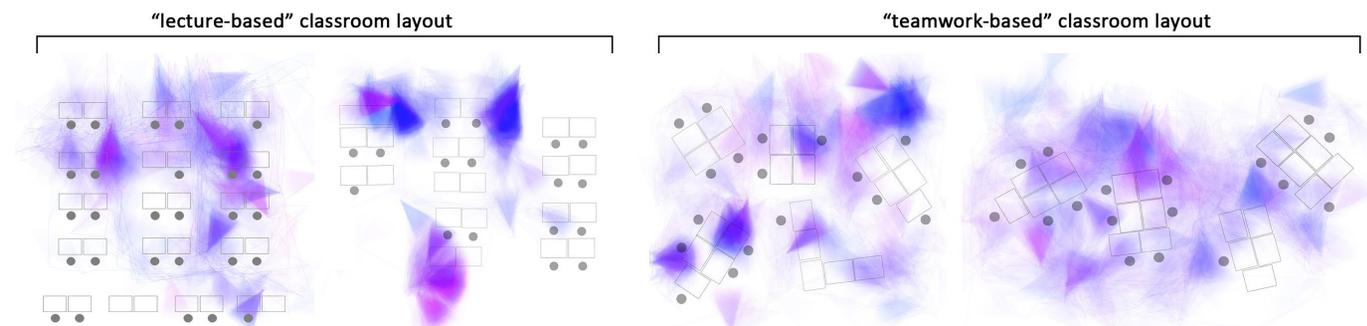

"lecture-based" classroom layout    "teamwork-based" classroom layout

**Figure 4**: Examples of how teachers' proxemics patterns might appear differently in two different types of classrooms.

## References

[1] An, P., Bakker, S., Ordanovski, S., Taconis, R., Paffen, C.L.E. and Eggen, B. 2019. Unobtrusively Enhancing Reflection-in-Action of Teachers through Spatially Distributed Ambient Information. *Proceedings of the 2019 CHI Conference on Human Factors in Computing Systems - CHI '19* (New York, NY, USA, 2019), 1–14.

[2] An, P., Ordanovski, S., Bakker, S., Taconis, R. and Eggen, B. 2018. ClassBeacons: Designing distributed visualization of teachers' physical proximity in the classroom. *TEI 2018 - Proceedings of the 12th International Conference on Tangible, Embedded, and Embodied Interaction* (New York, NY, USA, 2018), 357–367.

[3] Andrienko, G., Andrienko, N., Bak, P., Keim, D. and Wrobel, S. 2013. *Visual analytics of movement*. Springer Science & Business Media.

[4] Caldwell, J. 1979. Basic Techniques for Early Classroom Intervention. *Pointer*. 24, 1 (1979), 53–60.

[5] Chin, H.B., Mei, C.C.Y. and Taib, F. 2017. Instructional Proxemics and Its Impact on Classroom Teaching and Learning. *International Journal of Modern Languages and Applied Linguistics*. 1, 1 (2017).

[6] Dodge, S., Weibel, R. and Lautenschütz, A.-K. 2008. Towards a taxonomy of movement patterns. *Information visualization*. 7, 3–4 (2008), 240–252.

[7] Eriksen, J.B. 2015. Visualization of crowds from indoor positioning data. NTNU.

[8] Gunter, P.L., Shores, R.E., Jack, S.L., Rasmussen, S.K. and Flowers, J. 1995. On the Move Using Teacher/Student Proximity to Improve Students' Behavior. *TEACHING Exceptional Children*. 28, 1 (Sep. 1995), 12–14. DOI:https://doi.org/10.1177/004005999502800103.

[9] Hall, E.T. 1963. A System for the Notation of Proxemic Behavior. *American Anthropologist*. 65, 5 (Oct. 1963), 1003–1026. DOI:https://doi.org/10.1525/aa.1963.65.5.02a00020.

[10] Hyougo, Y., Misue, K. and Tanaka, J. 2014. Directional aggregate visualization of large scale movement data. *2014 18th International Conference on Information Visualisation* (2014), 196–201.

[11] in-door positioning sensor kit: *https://www.pozyx.io/*.

[12] Kale, U. 2008. Levels of interaction and proximity: Content analysis of video-based classroom cases. *The Internet and Higher Education*. 11, 2 (Jan. 2008), 119–128. DOI:https://doi.org/10.1016/j.iheduc.2008.06.004.

[13] Koren, H.-K.S. and Krogstie, J. 2016. Visualizing Large Indoor Positioning Data Sets in Web Browsers. (2016).

[14] Lanir, J., Kuflik, T., Sheidin, J., Yavin, N., Leiderman, K. and Segal, M. 2017. Visualizing museum visitors' behavior: Where do they go and what do they do there? *Personal and Ubiquitous Computing*. 21, 2 (2017), 313–326.

[15] Lim, F. V, O'Halloran, K.L. and Podlasov, A. 2012. Spatial pedagogy: mapping meanings in the use of classroom space. *Cambridge Journal of Education*. 42, 2 (Jun. 2012), 235–251. DOI:https://doi.org/10.1080/0305764X.2012.676629.

[16] Martinez-Maldonado, R. 2019. "I Spent More Time with that Team." *Proceedings of the 9th International Conference on Learning Analytics & Knowledge - LAK19* (New York, New York, USA, 2019), 21–25.

[17] Martinez-Maldonado, R., Echeverria, V., Santos, O.C., Santos, A.D.P. Dos and Yacef, K. 2018. Physical Learning Analytics: A Multimodal Perspective. *Proceedings of the 8th International Conference on Learning Analytics and Knowledge* (New York, NY, USA, 2018), 375–379.

[18] Martinez-Maldonado, R. and Nieto, G.F. 2019. Multimodal Analytics for Classroom Proxemics. *ALASI2019 - Australian Learning Analytics Summer Institute 2019* (2019).

[19] Men, L., Bryan-Kinns, N. and Bryce, L. 2019. Designing spaces to support collaborative creativity in shared virtual environments. *PeerJ Computer Science*. 5, (2019), e229.

[20] Shores, R.E., Gunter, P.L. and Jack, S.L. 2017. Classroom Management Strategies: Are They Setting Events for Coercion? *http://dx.doi.org/10.1177/019874299301800207*. (Feb. 2017). DOI:https://doi.org/10.1177/019874299301800207.

[21] Sills-Briegel, T.M. 1996. Teacher-Student Proximity and Interactions in a Computer Laboratory and Classroom. *The Clearing House: A Journal of Educational Strategies, Issues and Ideas*. 70, 1 (Oct. 1996), 21–23. DOI:https://doi.org/10.1080/00098655.1996.10114351.

[22] Yaeli, A., Bak, P., Feigenblat, G., Nadler, S., Roitman, H., Saadoun, G., Ship, H.J., Cohen, D., Fuchs, O. and Ofek-Koifman, S. 2014. Understanding customer behavior using indoor location analysis and visualization. *IBM Journal of Research and Development*. 58, 5/6 (2014), 1–3.